\documentstyle[11pt,newpasp,twoside,epsf]{article}
\begin{document}

\title{Coupling Chemical Evolution with SPH}

\author{Giovanni Carraro}
\affil{Astronomy Department, Padova University, Vicolo Osservatorio 2, I-35122, Padova}
\author{Laura Portinari}
\affil{Theoretical Astrophysics Center, Juliane Marie Vej 30, DK-2100, Copenhagen}
\author{Cesario Lia}
\affil{SISSA/ISAS, Via Beirut 2, I-34014, Trieste}

\begin{abstract}
We present a new statistical algorithm of  Chemical Evolution
implemented into the Tree-SPH code developed by Lia \& Carraro (2000).
\end{abstract}

\noindent
In Smoothed Particles Hydrodynamics (SPH) codes with a large number
of particles, star formation
as well as gas and metal restitution from dying stars can be treated
statistically. This approach allows
to include detailed chemical evolution and gas {\mbox{re-ejection}}
with minor
computational effort. Here we draw the attention on a new statistical
algorithm for star formation (SF) and chemical evolution, especially
conceived for SPH simulations with large numbers of particles,
and for parallel SPH codes described in full
detail elsewhere in Lia et al (2001).\\
Basically  SF is treated with a probabilistic approach,
and when stars are formed they are considered as a single stellar
population (SSP) for which gas restitution and metal production
for different species (He, Fe, Ca, Z, Mg, O, N, C, S and total
metallicity Z) are 
computed as a function of the age of the SSP,
taking into account the role of single stars of all
masses and lifetimes, as well as of Type Ia SN.
Our simulations include also Feed-Back from SN\ae~ of both type Ia and II,
and stellar winds. A diffusion mechanism is adopted to spread metals
in the interstellar medium (ISM).
A novelty of our approach is that also the gas and metals
restitution is treated in a probabilistic manner.
\\

\noindent
We present two astrophysical simulations
obtained with this algorithm.\\
In the first one, we follow the formation of  an individual disc--like
galaxy, predict the final structure, the metallicity evolution
and metallicity gradients.
In the left panel of Fig.~1 we present the trend of [O/Fe] vs. [Fe/H]
at the end of the simulation, and compare it with observational data
for the solar vicinity.\\

\noindent
In the second one we simulate the formation and evolution
of a cluster of galaxies, to demonstrate the capabilities
of the algorithm in investigating the chemo-dynamical evolution
of the intergalactic medium in a cosmological context.
The right panel of Fig.~1 shows the enrichment of the intra-cluster medium
in global metallicity (Z), [Fe/H] and [O/Fe]. Whenever it is possible,
a comparison is made with observational data for cluster of galaxies.
As one can note, the agreement is very good.\\

\noindent
Our implementation is intended to be easily implemented into any SPH code.
Anybody interested is welcome.\\

\acknowledgments
GC deeply thanks IAU for financial support. The authors acknowledge
useful discussions with Jesper Sommer-Larsen and Claudio Dalla Vecchia.

\begin{figure}
\plottwo{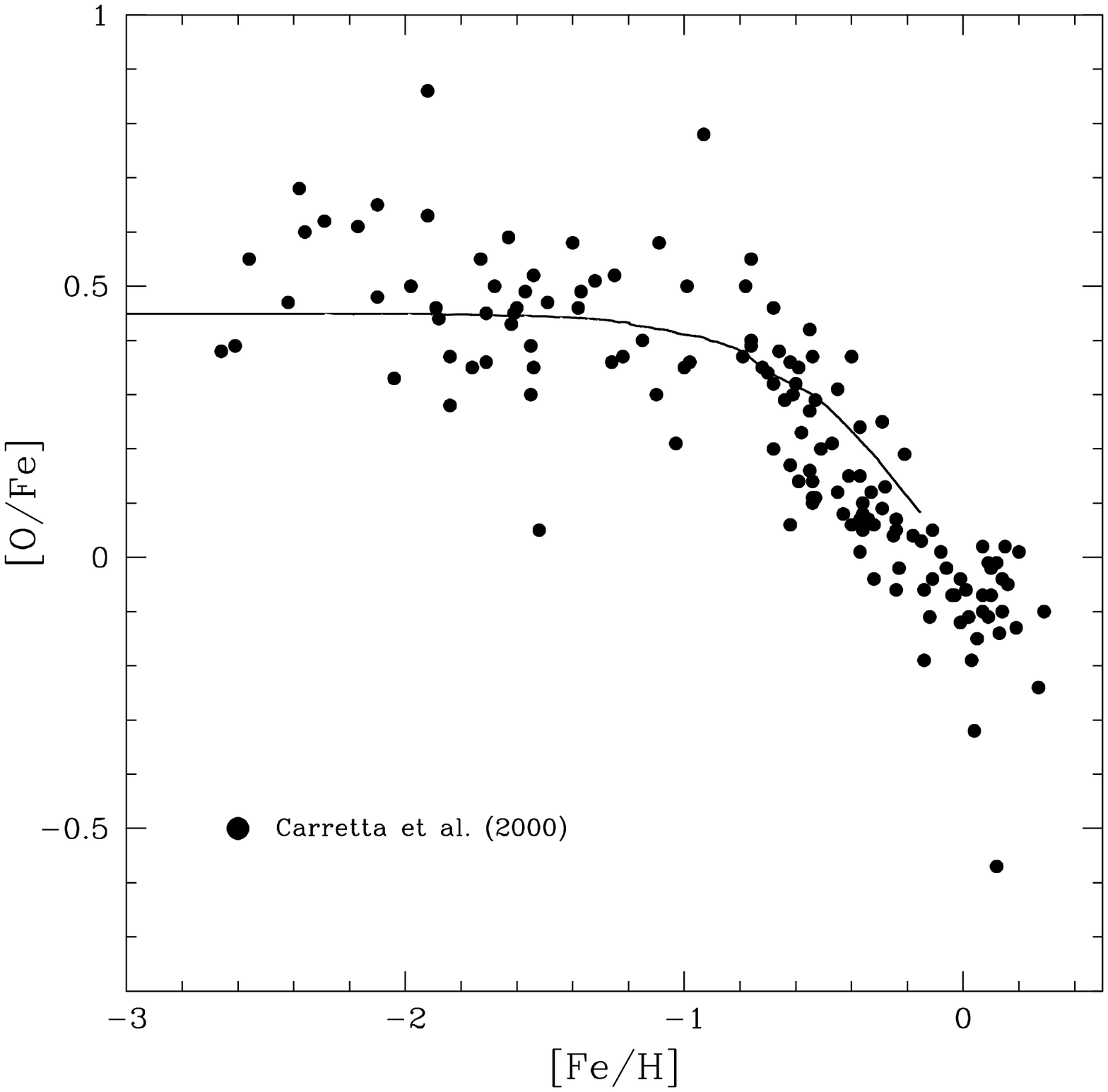}{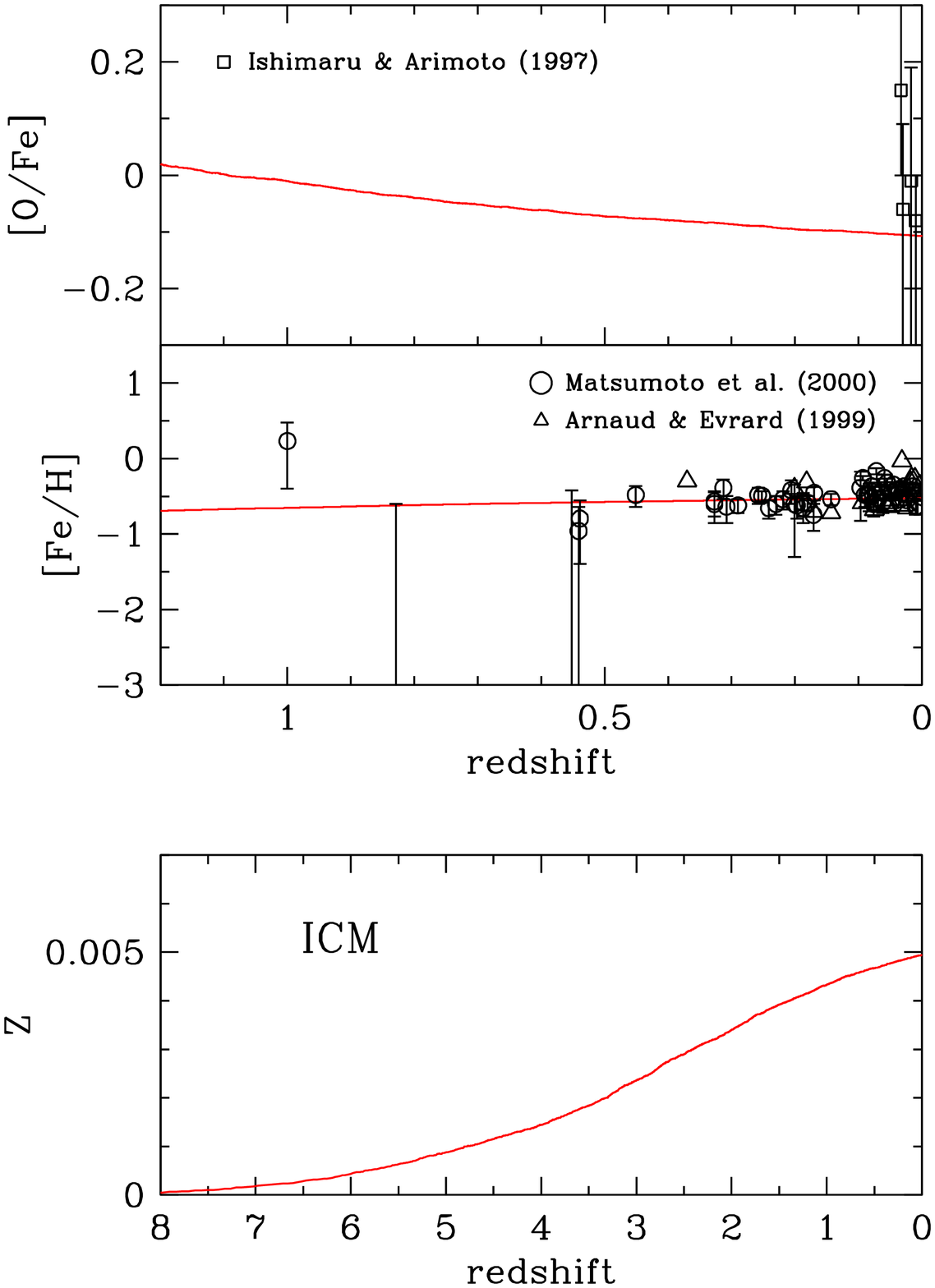}
\caption{Chemical enrichemnt of a disk galaxy model (left panel)
and of the intra-cluster medium (right panle).}
\end{figure}

\end{document}